\documentstyle[amssymb,prl,aps,psfig,floats]{revtex}
\twocolumn

\begin{document}
\draft

\wideabs{

\title{Universal flow diagram for the magnetoconductance in
disordered GaAs layers}
\author{S. S. Murzin$^{\text{1,2}}$, M. Weiss$^{\text{1}}$, A. G. M. Jansen$^{\text{1}}$ and K. Eberl$^{\text{3}}$}
\address{$^{\text{1}}$Grenoble High Magnetic Field Laboratory, Max-Planck-Institut
f\"{u}r Festk\"{o}rperforschung and\\ Centre National de la Recherche
Scientifique, BP 166, F-38042, Grenoble Cedex 9, France \\
$^{\text{2}}$Institute of Solid State Physics RAS, 142432, Chernogolovka,
Moscow District., Russia\\
$^{\text{3}}$Max-Planck-Institut f\"{u}r Festk\"{o}rperforschung, Postfach
800 665 D-70569, Stuttgart, Germany}
\maketitle

\begin{abstract}
The temperature driven flow lines of the diagonal and Hall
magnetoconductance data $(G_{xx},G_{xy})$ are studied in heavily
Si-doped, disordered GaAs layers with different thicknesses.
The flow lines are quantitatively well described  by a recent universal scaling
theory developed for the case of duality symmetry. The separatrix $G_{xy}=1$
(in units $e^{2}/h$) separates an insulating state
from a spin-degenerate quantum Hall effect (QHE) state.
The merging into the insulator or the QHE state at low temperatures happens
along a semicircle separatrix $G_{xx}^{2}+(G_{xy}-1)^{2}=1$ which is divided
by an unstable fixed point at $(G_{xx},G_{xy})=(1,1)$.
\end{abstract}


}

\narrowtext

In spite of considerable efforts in theoretical and experimental research on
the Quantum Hall Effect (QHE) for many years, the complete description of
its evolution for decreasing temperature is still unsatisfactory at the
moment. About 20 years ago a flow-diagram \cite{Khm} for the coupled
evolution of the diagonal ($G_{xx}$) and Hall ($G_{xy}$) conductivities was
sketched for increasing sample size $L$ (or, equivalently, increasing phase
breaking length $L_{\phi }$ for finite decreasing temperatures) on the basis
of a two-parameter scaling approach to the QHE \cite{Pr}. With increasing
system size ($L\rightarrow \infty$) $G_{xx}$ tends to zero ($%
G_{xx}\rightarrow 0$) while $G_{xy}$ becomes quantized ($G_{xy}\rightarrow i$%
, $i$ is an integer, $G_{xx}$ and $G_{xy}$ are in units $\ e^{2}/h$). The
points $(G_{xx}(L),G_{xy}(L))$ flow on lines merging into the QHE plateau
states characterized by one of the fixed points $(0,i)$. In addition there
are unstable fixed points in between these plateaus at $%
(G_{xx}^{c},G_{xy}^{c}=i+1/2)$ where the flow lines terminate, meaning that
points with $G_{xy}=i+1/2$ maintain their Hall conductance for all $L$.
Numerical calculations for a system of noninteracting electrons at high
magnetic fields in the lowest Landau level give $G_{xx}^{c}\approx 1/2$ in
the presence of different random potentials \cite{Huo}. For sufficiently low
temperatures the $(G_{xx},G_{xy})$ data flow on a separatrix in the flow
diagram which for a smooth disorder potential has been derived to be a
semicircle of the form \cite{RF}
\begin{equation}
G_{xx}^{2}+\left[ G_{xy}-(i+1/2)\right] ^{2}= 1/4,  \label{sl1}
\end{equation}
with $G_{xx}^{c}=1/2$ and $G_{xy}^{c}=i+1/2$.

Recently it has been shown \cite{BDD} that the semicircle law follows solely
from the consistency of the law of corresponding states, which was already
used to deduce the complete set of fractional and integer quantum Hall
states from the integer $i=1$ state \cite{KLZ}. Moreover, exact flow lines
for the integer and fractional QHE were derived \cite{Dolan} from duality
and particle-hole symmetries \cite{BD}, which underlie the law of
corresponding states. Their shape does not depend on details of the 2D
system. The separatrix are the above given semicircles and the vertical
lines $G_{xy}=i+1/2 $.

In the presented work we explore the temperature driven flow diagram of $%
G_{xx}(T)$ versus $G_{xy}(T)$ for disordered heavily Si-doped GaAs layers
with different thicknesses from 40 to 27~nm in a large temperature range
from 4.2~K down to 40~mK. At low temperatures these samples are situated in
the transition region between a QHE state and an insulating state. For the
temperature evolution a quantitative agreement is found with the universal
theory \cite{Dolan} for the flow lines of the $(G_{xx},G_{xy})$ data points.

\begin{figure}[t]
~~~\psfig{figure=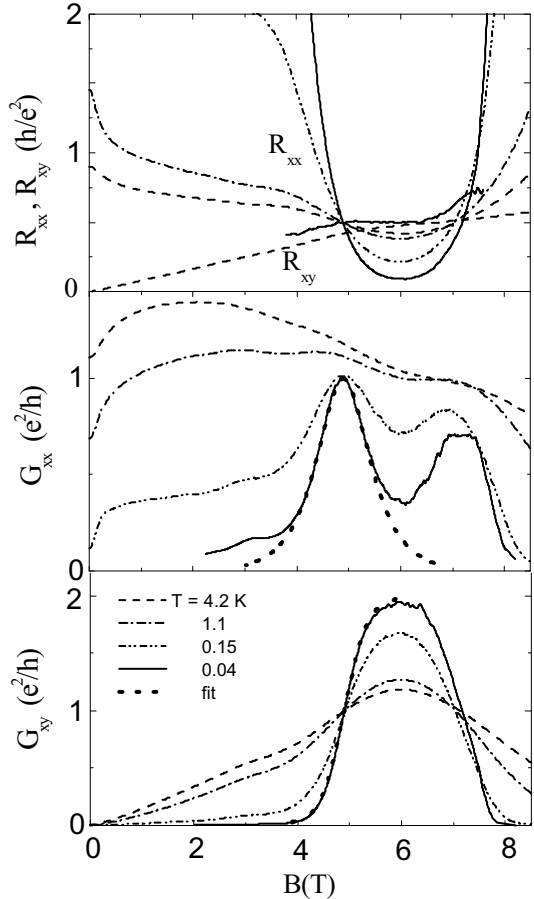,width=7cm}
\caption{Magnetic field dependence of the diagonal ($R_{xx}$, per square)
and Hall ($R_{xy}$) resistance and the diagonal ($G_{xx}$) and Hall ($G_{xy}$%
) conductance for sample~30 in a magnetic field perpendicular to the heavily
doped GaAs layer at different temperatures. Dotted lines for $G_{xx}$ and $%
G_{xy}$ show the result of a theoretical fit around $B_{c}=4.9$~T. }
\label{R30}
\end{figure}

The temperature evolution of the flow diagram was already studied in many
experiments \cite{flow}, but showed inconsistencies with respect to theory.
As an example, values reported for $G_{xx}^{c}$ in the literature range from
0.02 up to 0.8 (for the transition $i\rightarrow i+1$), and even different
values for different $i$ on the same sample have been observed. The
semicircle relation was observed in the dependence $G_{xx}(G_{xy})$ with $%
G_{xx}$ and $G_{xy}$ driven by magnetic field in a SiGe-Ge-SiGe quantum
well at rather high temperatures of 0.3-3 K. The position of the unstable
fixed point however was essentially different from $(1/2,1/2)$ \cite{Hilke}.
Although $G_{xx}^{c}\approx 1/2$ and $%
G_{xy}^{c}\approx i+1/2$ were observed at the transition from the insulator
to the quantum Hall state ($i=0$) in $\delta $-doped GaAs \cite{Hughes} and
between two QHE plateaus with $G_{xy}=1$ and $2$ ($i=1$) in a Si-Ge hole
system \cite{Dunford}, in most experiments these values differ significantly
from $1/2$ and $i+1/2$, respectively.

There are several reasons leading to a difference between theoretical and
experimentally observed critical values of $G_{xx}^{c}$ and $G_{xy}^{c}$,
like macroscopic inhomogeneities of the sample, enhanced current densities
near the edges of even homogeneous samples (this explains probably the low
values $G_{xx}^{c}\approx 0.02$ observed in high mobility GaAs
heterostructures) and spin effects.

The theory introduced in Ref. \cite{Dolan} has been developed for spinless
(or totally spin polarized) electrons. Therefore, the most favorable
candidate for an experimental study of the flow diagram under integer QHE
conditions is a disordered system with a small $g$-factor such that the
spin-splitting $g\mu _{B}B$ ($\mu _{B}$ is the Bohr magneton) is small with
respect to the disorder broadening and will only show up in the flow diagram
at rather low temperatures \cite{Khm3}. As we have shown in previous
investigations on similar samples, electron-electron interaction cannot be
neglected in the systems which are subject of the present work \cite{MWJK}.
For $G_{xx}>1$ interaction mainly leads to temperature dependent flow of the
$(G_{xx}(T),G_{xy}(T))$ data \cite{m98} and a dependence of the localization
length on interaction \cite{MWJK}. For $k_{B}T\ll \mu _{B}gB$ ($k_{B}$ is
the Boltzmann constant) only interaction
of electrons with the same spin leads to a renormalization of the
conductance \cite{Fin}. Therefore, under these conditions and in the absence
of spin-flip scattering, electrons with different spin can be considered as
two independent, totally spin polarized electron systems. For such a
situation, one should substitute $G_{ij}$ by $G_{ij}/2$ in the above given
expressions for the conductances leading to a semicircle relation of the
form
\begin{equation}
G_{xx}^{2}+(G_{xy}-(2i+1))^{2}=1 \label{sl2}
\end{equation}
with the vertical separatrix $G_{xy}=2i+1$.

The disordered GaAs samples were prepared by molecular-beam epitaxy. On a
GaAs (100) substrate were successively grown an undoped GaAs layer (0.1~$\mu
$m), a periodic structure of 30~$\times $ GaAs/AlGaAs(10/10 nm), an undoped
GaAs layer (0.5$\mu $m), the heavily Si-doped GaAs of a nominal thickness of
$d=27,30,34$ and $40$~nm and a Si-donor concentration of $1.5\times 10^{17}$
cm$^{-3}$, followed by an cap layer of 0.5 $\mu $m undoped GaAs. The number
given for a sample corresponds to the thickness of its doped layer. Hall bar
geometries of width 0.2~mm and length 2.8~mm were etched out of the wafers.
A phase sensitive ac-technique was used for the magnetotransport
measurements down to 40~mK with the applied magnetic field up to 12~T
perpendicular to the layers. For samples 27 and 30 the absolute values of
the Hall resistance $R_{xy}$ were about $10\%$ different for two opposite
directions of the magnetic field. The average has been taken as $R_{xy}$.
The electron densities per square as derived from the slope of the Hall
resistance $R_{xy}$ in weak magnetic fields ($0.5-3$~T) at $T=4.2$~K are $%
N_{s}=3.7,4,5.5$ and $6.2\times 10^{11}$ cm$^{-2}$ for samples 27, 30, 34
and 40, respectively. The ''bare'' high temperature mobilities $\mu _{0}$
are about $1300,1400,1900$ and $2300$~cm$^{2}$/Vs. Because of the rather
large quantum corrections to the conductivity, even in zero magnetic field
at 4.2 K, we used the approximate relation $\mu _{0}=R_{xy}/BR_{xx}$ in the
intersection point of the $R_{xx}(B)$ curves for different temperatures.
Previous experimental studies of the flow diagram have been performed on
much purer samples with at least an order of magnitude higher mobility.

\begin{figure}[t]
~~~\psfig{figure=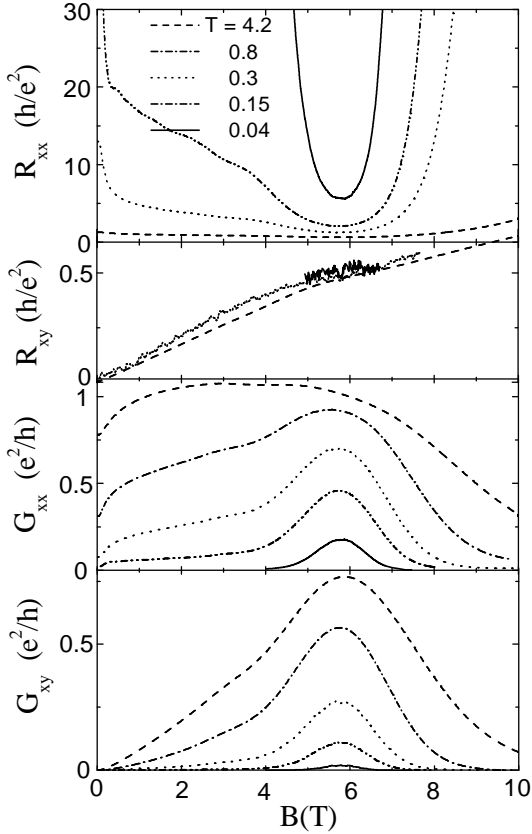,width=7cm}
\caption{Magnetic field dependence of the diagonal ($R_{xx}$, per square) and
Hall ($R_{xy}$) resistance, diagonal ($G_{xx}$) and Hall ($G_{xy}$) conductance for
sample~27 in a magnetic field perpendicular to the heavily doped GaAs layer
at different temperatures. }
\label{R27}
\end{figure}

Samples $34$ and $40$ reveal a wide QHE plateau from $\approx 6$ up to
$\approx 11$~T with the value $R_{xy}=h/2e^{2}$
(i.e. $i=2$ for a spin degenerate lowest Landau-level
occupation) accompanied by an exponentially small value of $R_{xx}$ at low
temperatures $T\lesssim 0.3$ K. The magnetoresistance data of sample 40 are
presented in Ref.\cite{MWJK}.

In Fig.\ref{R30} the magnetotransport data of the diagonal ($R_{xx}$, per
square) and Hall ($R_{xy}$) resistance (both given in units of $h/e^{2}$), and
of the diagonal ($G_{xx}$) and Hall ($G_{xy}$) conductance have been
plotted for sample~30. At $T=4.2$~K, $R_{xx}$ depends on magnetic field
rather weakly and has only a weak minimum at $B=6$~T, and $R_{xy}$ increases
linearly up to 5 T with a slightly smaller slope at higher fields. Such a
behaviour is typical for bulk samples in the extreme quantum limit. At the
lowest temperatures, the layer is insulating ($R\geq 100$) in zero magnetic
field. At low magnetic fields up to 0.5~T, the diagonal resistance $R_{xx}$
drops abruptly and continues to decreases more slowly between 0.5 and 4~T.
For fields between 5 and 7~T, a minimum is observed with a QHE plateau
in the Hall resistance with $R_{xy}=1/2$. The same QHE structure around 6~T
can be observed in the plotted conductance data. In the minimum of $G_{xx}$
near $B=6$ T the Hall conductance $G_{xy}$ increases from a value higher
than 1 (at 4.2~K) towards 2 at the lowest temperatures. The curves $%
G_{xy}(B) $\ for different temperature cross at one point with $G_{xy}=1$ at
$B_{c}=4.9 $~T. The diagonal conductivity tends towards 1 for decreasing
temperature at this critical field. In this critical point the derivative $%
dG_{xy}/dB$ shows a power law dependence $\sim T^{-\mu }$ with $\mu =0.48\pm
0.05$. For curves at $T>0.2$\ K there is a second crossing point at $B=7$
T with $G_{xy}\approx 1$. The second peak in $G_{xx}(B)$ has an amplitude
smaller than $e^{2}/h$ and is broader than the first one. We believe that
this second peak structure is a manifestation of spin-splitting.

In Fig.\ref{R27} the magnetotransport data have been plotted for sample~27.
At $T=4.2$ K these data are
similar to the data for sample 30. However, sample 27 shows insulating
behavior ($R_{xx}$ increases with decreasing temperature) at all magnetic
fields with a rather deep and narrow minimum in the field dependence $%
R_{xx}(B)$ at low temperatures. Note, that at the lowest temperature we can
measure $R_{xy}$ only near the minimum of $R_{xx}$ since outside this region
$R_{xy}\ll R_{xx}$. Within our accuracy  the sample reveals a
Hall-insulator state ($R_{xy}=0.5$, \cite{KLZ}) in this region. $G_{xx}$ and
$G_{xy}$ have peaks at $B\approx 6$~T.

The QHE in sample 30 is much less pronounced than in samples 34 and 40
due to the fact that the maximum of the high temperature Hall conductance
$G_{xy}^{0}(B)$ has a value of $\approx 1.2$ close to $1$ (see Fig.1).
For $G_{xy}^{0}\rightarrow 1$ the localization length diverges, and the system
is in the dissipative, non-quantized state.
For samples 34 and 40 with a maximum of $G_{xy}^{0}(B)$ close to 2,
quantization at $G_{xy}=2$ develops already at higher temperatures.
Although insulating for all fields, sample 27 shows a minimum in $R_{xx}$
and a maximum in $G_{xx}$ due to the proximity of $G_{xy}^{0}(B)$ to $1$
on the insulator side, giving a large localization length at its maximum.

\begin{figure}[t]
~~~\psfig{figure=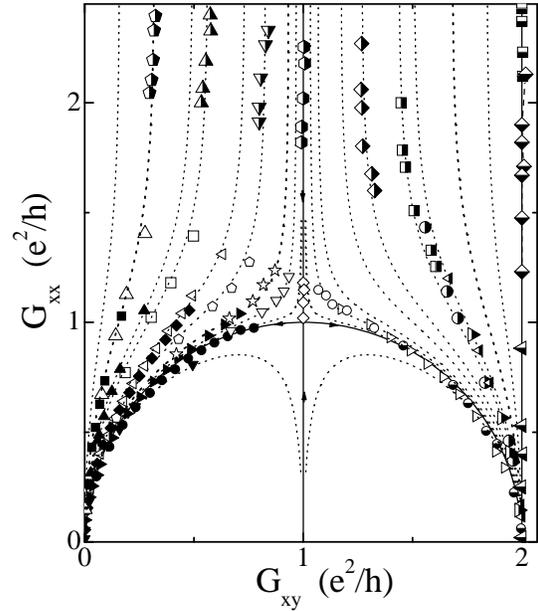,width=7cm}
\caption{Flow-diagram of the ($G_{xx}(T),G_{xy}(T)$) data points for the
investigated heavily doped GaAs layers with different thickness (filled
symbols are for sample 27, open - sample 30, half right filled - sample 34
and half bottom filled - sample 40) at different magnetic field values
(different type of symbols). Dotted lines show the theoretical flow lines.
Solid lines display the separatrix.}
\label{flow}
\end{figure}

In Fig.\ref{flow} the temperature evolution of the points ($%
G_{xx}(T),G_{xy}(T))$ has been plotted for the different samples at different
magnetic fields taken from 4.2 down to temperatures between 0.04 and 0.1~K
except for the flow lines of
sample 34 in weak magnetic fields ($1.4-2.4$~T) which start only at 1.1~K.
The data points at the lowest temperatures approach and, subsequently,
follow the semicircle dependence given in Eq.(\ref{sl2}). Their final low-temperature
limiting value depends on the initial high temperature Hall conductance $%
G_{xy}^{0}$ with respect to $G_{xy}=1$. Data points starting on the
semicircle follow this semicircle. The points starting for high temperatures
at $G_{xy}=1$ terminate at the lowest temperatures very close to $%
(G_{xx}^{c},G_{xy}^{c})=(1,1)$. The presented data on disordered GaAs layers
follow the trends expected from universal scaling arguments.

In the following we will give a quantitative estimate for the temperature
dependent evolution of the flow lines at constant magnetic field. The dotted
flow lines in Fig.\ref{flow} are plotted as a result of a numerical solution
of the equation $\arg (f)=\alpha $ for various $\alpha $, where
\begin{equation}
f=-\vartheta _{3}\vartheta _{4}/\vartheta _{2},
\end{equation}
with the Jacobi $\vartheta $ functions
\begin{eqnarray}
\vartheta _{2}(q) &=&2\sum_{n=0}^{\infty }q^{\left( n+1/2\right) ^{2}}\text{
, \ \ \ }\vartheta _{3}(q)=\sum_{n=-\infty }^{\infty }q^{n^{2}}\text{ ,}
\nonumber \\
\vartheta _{4}(q) &=&\sum_{n=-\infty }^{\infty }(-1)^{n}q^{n^{2}},
\end{eqnarray}
for $q=\exp \left[ i\pi (G_{xy}+iG_{xx})/2\right] $ \cite{Dolan}. The value $%
\alpha $ corresponds to the Hall conductance $G_{xy}^{\infty}$
for large $G_{xx}$ where the flow lines are vertical, with
$\alpha =\pi (1-G_{xy}^{\infty})$ (for the flow lines
above the semicircle Eq.(\ref{sl2})). The theoretical
flow-lines are in very good agreement with the experimental data and are
universally determined by the limiting $G_{xy}^{\infty}$ values.

The rate of flow is determined by the temperature dependence of $G_{xx}(T)$
via the parameter $s-s_{0}=\ln (f/f_{0})$ where $f_{0}=f(s_{0})$. For flow
along the semicircle from the critical point at ($1,1$), where $f_{0}=1/4$, we have $%
s=\ln (4f)$. In this case \cite{Dolan}
\begin{eqnarray}
G_{xx} &=&2\frac{K^{\prime }(w)K(w)}{K(w)^{2}+K^{\prime }(w)^{2}}\text{ \ \
\ \ }G_{xy}=2\frac{K^{\prime }(w)^{2}}{K(w)^{2}+K^{\prime }(w)^{2}}
\nonumber \\
w &=&\sqrt{\frac{1\pm \sqrt{1-\exp (-s)}}{2}}  \label{GG}
\end{eqnarray}
where $K(w)$ is the complete elliptic function of the second
kind, with $K^{\prime }(w)\equiv K(\sqrt{1-w^{2}})$. The temperature
dependence of $G_{xx}$ and $G_{xy}$ along the semicircle for
samples 27, 30 and 40 can be fitted by Eq.(\ref{GG}) and $s=c/T^{p}$
(with $T$ in K) with $p=1.16, 0.94, 1.1\pm
0.1$ and $c=0.83, 0.58, 3.5$, correspondingly.
For sample 30 the value of $p$ is two times larger than the value of
$\mu =0.48\pm 0.05$ extracted from the temperature dependence of
$dG_{xy}/dT$, in accordance with the proposed dependence
$s \propto \left( \Delta B/T^{\mu }\right) ^{2}$ \cite{Dolan} ($\mu$
is the critical exponent and $\Delta B=B-B_{c}$).
As shown in Fig. \ref{R30} the data
for $G_{xx}(B)$ and $G_{xy}(B)$ of sample 30 at the lowest
temperature are well described by Eq.(\ref{GG})
and $s=22\left( \Delta B\right) ^{2}$ (with $\Delta B$ in T)
around the critical point $B_{c}=4.9$ T.

For large values of $G_{xx}$ its temperature dependence is mostly due to
electron-electron interaction and the parameter $s$ can be written as $%
s=(\lambda /\pi )\ln L_{T}$, where $L_{T}$ is the coherence length for
interaction and the constant of interaction $\lambda $ is a material
parameter which depends on magnetic field ($\lambda <1$). Note that compared
to the two-parameter scaling theory $L_{\phi }$ is replaced by$L_{T}$. The
interaction effects accelerate the motion along the lines.

In summary, the flow-diagram of ($G_{xx}(T),G_{xy}(T)$) data for strongly
disordered GaAs layers is well described by the universal expressions
following from duality and particle-hole symmetries. Electron-electron
interaction leads to an accelerated flow rate but does not change the shape
of the flow lines.

This work is supported by RFBR and INTAS. We would like to thank B. Lemke
for her help in the preparation of the samples.

\end{document}